\documentclass[aps,pre,twocolumn,groupedaddress,showpacs,showkeys]{revtex4}

\usepackage{graphicx}
\usepackage{dcolumn}
\usepackage{bm}
\usepackage{amssymb}
\usepackage{amsmath}

\begin{document}


\title{Generalized synchronization: a modified system approach\footnote{This
  paper has been published in Physical Review E (Statistical,
  Nonlinear, and Soft Matter
  Physics) {\bf71}, 6 (2005) 067201}}


\author{Alexander~E.~Hramov}
\email{aeh@cas.ssu.runnet.ru}
\author{Alexey~A.~Koronovskii}
\email{alkor@cas.ssu.runnet.ru}
\affiliation{Faculty of Nonlinear
Processes, Saratov State University, Astrakhanskaya, 83, Saratov,
410012, Russia}



\date{\today}

\begin{abstract}
The universal mechanism resulting in the generalized
synchronization regime arising in the chaotic oscillators with the
dissipative coupling has been described. The reasons of the
generalized synchronization occurrence may be clarified by means
of a modified system approach. The main results are illustrated by
unidirectionally coupled R\"ossler systems, R\"ossler and Lorenz
systems and logistic maps.
\end{abstract}


\maketitle




Chaotic synchronization is one of the fundamental phenomena,
widely studied recently~\cite{Pikovsky:2002_SynhroBook}, having
both theoretical and applied significance (e.g., for information
transmission by means of deterministic chaotic signals
\cite{Murali:1993_SignalTransmission,Chua:1997_Criptography}, in
biological 
and physiological~\cite{Glass:2001_SynchroBio} tasks, etc.).
Several different types of chaotic synchronization of coupled
oscillators, i.e. \emph{generalized synchronization} (GS)
\cite{Paoli:1989_ScalingBehavior,Rulkov:1995_GeneralSynchro},
\emph{phase synchronization} (PS) \cite{Pikovsky:2002_SynhroBook},
\emph{lag synchronization} (LS) \cite{Rosenblum:1997_LagSynchro}
and \emph{complete synchronization} (CS)
\cite{Pecora:1990_ChaosSynchro} are well known. There are also
attempts to find unifying framework for chaotic synchronization
of coupled dynamical systems \cite{Brown:2000_ChaosSynchro,%
Boccaletti:2002_SynchroPhysReport,Boccaletti:2001_UnifingSynchro,%
Hramov:2004_Chaos}.

One of the interesting and intricate types of the synchronous
behavior of unidirectionally coupled chaotic oscillators is the
generalized synchronization. The presence of GS between the
response $\mathbf{x}_{r}(t)$ and drive $\mathbf{x}_{d}(t)$ chaotic
systems means that there is some functional relation
${\mathbf{x}_r(t)=\mathbf{F}[\mathbf{x}_d(t)]}$ between system
states after the transient finished. This functional relation
$\mathbf{F}[\cdot]$ may be smooth or fractal. According to the
properties of this relation, GS may be divided into the strong
synchronization and week synchronization,
respectively~\cite{Pyragas:1996_WeakAndStrongSynchro}. There are
several methods to detect the presence of GS between chaotic
oscillators, such as the auxiliary system
approach~\cite{Rulkov:1996_AuxiliarySystem} or the method of
nearest neighbors~\cite{Rulkov:1995_GeneralSynchro,%
Pecora:1995_statistics}. It is also possible to calculate the
\emph{conditional Lyapunov exponents} (CLEs)~\cite{Pecora:1991_ChaosSynchro,%
Pyragas:1996_WeakAndStrongSynchro} to detect GS. The regimes of LS
and CS are also the particular cases of GS.

This paper aims to explain GS arising. We show the physical
reasons leading to GS appearance in unidirectionally coupled
chaotic systems. The causes of the generalized synchronization
arising may be clarified by means of a modified system approach.

Let us consider the behavior of two unidirectionally coupled
chaotic oscillators
\begin{equation}
\begin{split}
&\mathbf{\dot x}_d=\mathbf{H}(\mathbf{x}_d,\mathbf{g}_d)\\
&\mathbf{\dot x}_r=\mathbf{G}(\mathbf{x}_r,\mathbf{g}_r)+
\varepsilon\mathbf{P}(\mathbf{x}_d,\mathbf{x}_r),\\
\end{split}
\label{eq:Oscillators}
\end{equation}
where $\mathbf{x}_{d,r}$ are the state vectors of the drive and
response systems, respectively; $\mathbf{H}$ and $\mathbf{G}$
define the vector fields of these systems, $\mathbf{g}_d$ and
$\mathbf{g}_r$ are the controlling parameter vectors, $\mathbf{P}$
denotes the coupling term and $\varepsilon$ is the scalar coupling
parameter. If the dimensions of the drive and response systems are
$N_d$ and $N_r$ respectively, the behavior of the unidirectionally
coupled oscillators~(\ref{eq:Oscillators}) is characterized by the
\emph{Lyapunov exponent} (LEs) \emph{spectrum}
${\lambda_1\geq\lambda_2\geq\dots\geq\lambda_{N_d+N_r}}$. Due to
the independence of the drive system dynamics on the behavior of
the response one, the Lyapunov exponent spectrum may be divided
into two parts:  LEs of the drive system
${\lambda^d_1\geq\dots\geq\lambda^d_{N_d}}$ and the
CLEs~\cite{Pecora:1991_ChaosSynchro,Pyragas:1997_CLEsFromTimeSeries}
${\lambda^r_1\geq\dots\geq\lambda^r_{N_r}}$. The condition of GS
is $\lambda^r_1<0$ (see~\cite{Pyragas:1996_WeakAndStrongSynchro}
for detail).

The GS manifestation is mostly considered for two identical
systems with equal or mismatched parameters and diffusive type of
unidirectional coupling. Therefore, let us consider such systems
first, while the case of different systems and others coupling
types will be briefly discussed later. In the case of identical
systems the dimensions of the drive and response oscillators are
equal ($N_d=N_r=N$) and the equations~(\ref{eq:Oscillators}) may
be rewritten as
\begin{equation}
\begin{split}
&\mathbf{\dot x}_d=\mathbf{H}(\mathbf{x}_d,\mathbf{g}_d)\\
&\mathbf{\dot x}_r=\mathbf{H}(\mathbf{x}_r,\mathbf{g}_r)+
\varepsilon\mathbf{A}(\mathbf{x}_d-\mathbf{x}_r),
\end{split}
\label{eq:Oscillators1}
\end{equation}
where $\mathbf{A}={\{\delta_{ij}\}}$ is the coupling matrix,
$\delta_{ii}=0$ or $1$ and $\delta_{ij}=0$ ($i\neq j$). It is
clear, that the dynamics of the response system may be considered
as the non-autonomous dynamics of the \emph{modified system}
\begin{equation}
\mathbf{\dot
x}_m=\mathbf{H}'(\mathbf{x}_m,\mathbf{g}_r,\varepsilon)
\label{eq:RsOsc}
\end{equation}
under the external force $\varepsilon\mathbf{A}\mathbf{x}_d$
\begin{equation}
\mathbf{\dot
x}_m=\mathbf{H}'(\mathbf{x}_m,\mathbf{g}_r,\varepsilon)+
\varepsilon\mathbf{A}\mathbf{x}_d, \label{eq:RsOsc&Force}
\end{equation}
where $\mathbf{H}'(\mathbf{x},\mathbf{g})=
\mathbf{H}(\mathbf{x},\mathbf{g})-\varepsilon\mathbf{A}\mathbf{x}$.
Note, that the term $-\varepsilon\mathbf{A}\mathbf{x}$ brings the
dissipation into the modified system~(\ref{eq:RsOsc}).

So, GS arising in~(\ref{eq:Oscillators1}) with parameter
$\varepsilon$ increasing may be considered as a result of two
cooperative processes taking place simultaneously. The first of
them is the growth of the dissipation in the
system~(\ref{eq:RsOsc}) and the second one is the increasing of
the amplitude of the external signal. Both processes are
correlated with each other by means of the parameter $\varepsilon$
and can not be realized in the coupled oscillators
system~(\ref{eq:Oscillators1}) independently. Nevertheless, let us
consider these processes separately to understand better the
mechanisms of GS arising. We start our considering with the
autonomous dynamics of the modified system~(\ref{eq:RsOsc}).

For this modified system, $\mathbf{x}_m(t)$, the quantity
$\varepsilon$ is the dissipation parameter. When $\varepsilon$ is
equal to zero the dynamics of the modified system
$\mathbf{x}_m(t)$ coincides with the response system
$\mathbf{x}_r(t)$ without coupling. With increasing of the
dissipation parameter $\varepsilon$ the dynamics of the modified
system (\ref{eq:RsOsc}) should be simplified. Therefore, the
system $\mathbf{x}_m(t)$ has to undergo a transition from chaotic
oscillations to periodic ones, and, perhaps, to the stationary
state (if the dissipation is large enough). In this case one of
the Lyapunov exponent, $\lambda_0^m$, of the modified system is
equal to zero (or negative if the stationary state takes place),
all other Lyapunov exponents are negative
(${0>\lambda^m_1\geq\dots\geq\lambda^m_{N-1}}$). It is important
to note, that the Lyapunov exponent spectrum of the modified
system~(\ref{eq:RsOsc}) differs from CLE spectrum
${\lambda^r_1\geq\dots\geq\lambda^r_{N}}$ of the
system~(\ref{eq:Oscillators1}), as CLEs depend on the drive system
dynamics in contrast to Lyapunov exponents of the modified system
$\mathbf{x}_m(t)$. Therefore, nobody can draw a conclusion about
the appearance of GS in the coupled oscillators
system~(\ref{eq:Oscillators1}) taking into account only Lyapunov
exponents of the modified system~(\ref{eq:RsOsc}).

On the contrary, the external signal in~(\ref{eq:RsOsc&Force})
tends to impose the dynamics of the drive chaotic oscillator
$\mathbf{x}_d(t)$ on the modified system $\mathbf{x}_m(t)$, and,
correspondingly, complicate its dynamics. Obviously, GS may take
place only if proper chaotic dynamics of the system
$\mathbf{x}_m(t)$ is suppressed by the dissipation. Only under
this condition the current state $\mathbf{x}_m(t)$ of the modified
system will be determined completely by the external signal, i.e.
$\mathbf{x}_m(t)=\mathbf{F}[\mathbf{x}_d(t)]$. According to the
equation~(\ref{eq:RsOsc&Force}), the functional relation
$\mathbf{x}_r(t)=\mathbf{F}[\mathbf{x}_d(t)]$ between the response
and drive systems will also take place, and, therefore, GS will be
observed.

So, GS arising  in the system~(\ref{eq:Oscillators1}) is possible
for such values of $\varepsilon$ parameter when the modified
system $\mathbf{x}_m(t)$ demonstrates the periodic oscillations or
the stationary state. It is well known, that even the harmonic
external signal can cause the chaotic oscillations in the
dynamical system with periodical dynamics. Therefore, the periodic
regime should be stable enough for the external force not to
excite proper chaotic dynamics of the modified system. So, the
difference between  the parameter values $\varepsilon_{p}$ when
the periodic oscillations take place in the
system~(\ref{eq:RsOsc}) and $\varepsilon_{GS}$ when GS in the
system~(\ref{eq:Oscillators1}) can be observed has to be large
enough. At the same time, the amplitude of the external signal is
small enough in comparison with the amplitude of periodic
oscillations in the modified system $\mathbf{x}_m(t)$ (in the case
when the periodical regime takes place in~(\ref{eq:RsOsc})). So,
the generalized synchronization looks like the week chaotic
excitation of the periodical motion.

The similar conclusion is also correct for the stationary regime
of the system $\mathbf{x}_m(t)$ when GS manifests itself as
chaotic perturbation of the fixed state. The system dynamics can
be considered as transient converging to the ``fixed'' point
moving under the external force in the phase space of the modified
system~(\ref{eq:RsOsc}). Let us suppose now that controlling
parameters $\mathbf{g}_{r,d}$ of the considered response and drive
systems differ from each other slightly and the value of parameter
$\varepsilon$ is large enough. In this case the transient is very
short and the state of the modified system follows the perturbed
``fixed'' state essentially small time $\tau$ of delay, therefore
the regime of LS can be observed.

Let us consider several examples of GS to illustrate the concept
described above. As the first system we have selected two
unidirectionally coupled R\"ossler oscillators
\begin{equation}
\begin{array}{l}
\dot x_{d}=-\omega_{d}y_{d}-z_{d},\\
\dot y_{d}=\omega_{d}x_{d}+ay_{d},\\
\dot z_{d}=p+z_{d}(x_{d}-c),\\
\\
\dot x_{r}=-\omega_{r}y_{r}-z_{r} +\varepsilon(x_{d}-x_{r}),\\
\dot y_{r}=\omega_{r}x_{r}+ay_{r},\\
\dot z_{r}=p+z_{r}(x_{r}-c),
\end{array}
\label{eq:Roesslers}
\end{equation}
where $\varepsilon$ is a coupling parameter, $\omega_r=0.95$. The
control parameter values have been selected by analogy
with~\cite{Zhigang:2000_GSversusPS} as $a=0.15$, $p=0.2$,
$c=10.0$. Correspondingly, the modified R\"ossler system is
\begin{equation}
\begin{array}{l}
\dot x_{m}=-\omega_{r}y_{m}-z_{m} -\varepsilon x_{m},\\
\dot y_{m}=\omega_{r}x_{m}+ay_{m},\\
\dot z_{m}=p+z_{m}(x_{m}-c).
\end{array}
\label{eq:ModRoesslers}
\end{equation}
In Fig~1,\textit{a} the bifurcation diagram for the
system~(\ref{eq:ModRoesslers}) is shown. It is clear that this
system undergoes the transition from chaotic to periodic
oscillations through the inverse cascade of period doubling. The
dependence of two largest Lyapunov exponents $\lambda^m_{0,1}$ on
the parameter $\varepsilon$ is presented in Fig.~1,\textit{b}. One
can easily see, that starting from the value $\varepsilon_p\approx
0.06$ the periodic oscillations take place in the modified
system~(\ref{eq:ModRoesslers}).

\begin{figure}[t]
\centerline{\includegraphics*[scale=0.5]{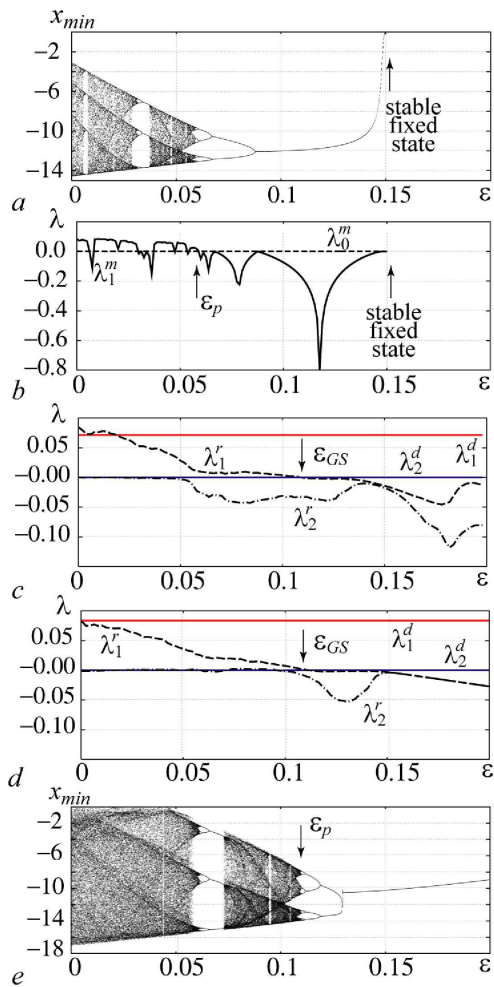}} \caption{The
bifurcation diagram (\textit{a}) and the dependence of two
Lyapunov exponent $\lambda^m_{0,1}$ (\textit{b}) of the modified
R\"ossler system~(\ref{eq:ModRoesslers}) on the parameter
$\varepsilon$. The third Lyapunov exponent is about
$\lambda^m_{2}\approx -9.7$ and is not significant for our
considering. The value of parameter $\varepsilon_p$ when the
modified system starts demonstrating the periodic dynamics is
shown by an arrow. (\textit{c,d}) The dependence of the Lyapunov
exponent spectrum on the parameter $\varepsilon$ for slightly
($\omega_d=0.99$) and greatly ($\omega_d=1.3$) detuned R\"ossler
systems, respectively. The onset of GS is marked by an arrow.
Conditional Lyapunov exponents are presented by dashed
($\lambda^r_1$) and dotted ($\lambda^r_2$) lines. (\textit{e}) The
bifurcation diagram for the non-autonomous modified system. The
first equation in~(\ref{eq:ModRoesslers}) is replaced by ${\dot
x_{m}=-\omega_{r}y_{m}-z_{m} -\varepsilon x_{m}+A\cos(\Omega t)}$
where $A=1.32$, $\Omega=1.0$ that simulates the parameters of the
drive R\"ossler system. The value of parameter $\varepsilon_p$
corresponding to the onset of the periodic oscillation is shown by
an arrow}
\end{figure}

Fig~1,\textit{c} demonstrates the dependence of the fourth largest
Lyapunov exponents of coupled R\"ossler
oscillators~(\ref{eq:Roesslers}) with the slight mistuning of the
control parameter $\omega_d$ (${\omega_d=0.99}$) on the coupling
strength $\varepsilon$. Two of them, $\lambda^d_1$ and
$\lambda^d_2$ correspond to the behavior of the drive system,
therefore they do not depend on $\varepsilon$. Two other
quantities $\lambda_{1,2}^r$ are the conditional Lyapunov
exponents. When the coupling parameter $\varepsilon$ is equal to
zero, $\lambda^r_1>0$ and $\lambda^r_2=0$. With parameter
$\varepsilon$ increasing the second CLE $\lambda_2^r$ becomes
negative ($\varepsilon\approx 0.04$), but the dynamics of the
modified system~(\ref{eq:ModRoesslers}) remains still chaotic
($\lambda^m_1>0$). With further increasing of $\varepsilon$ value
the dynamics of the modified system~(\ref{eq:ModRoesslers})
becomes periodical (see Fig.~1,\textit{a,b}), but GS is not yet
observed. It takes place only when the periodical regime of the
modified system~(\ref{eq:ModRoesslers}) is stable enough
($\varepsilon_{GS}\approx 0.11$). In this case the period-1 cycle
is realized in the modified R\"ossler system. Below the critical
value $\varepsilon_c\approx 0.15$ the modified R\"ossler system
comes to the stationary state. Note, that when the periodical
regimes take place in the modified system, the value of the
highest CLE is slightly negative if GS is realized. As soon as the
stationary state of the modified system~(\ref{eq:ModRoesslers})
becomes stable the value of $\lambda_1^m$ starts to decrease
rapidly.

Note also, that the onset of GS is determined by the stability of
the periodic regimes of the modified system~(\ref{eq:RsOsc}) which
does not depend on the mismatch of the control parameters
$\mathbf{g}_{d,r}$ of the unidirectionally coupled oscillators.
The stability of the periodical regimes is caused by the property
of the modified system only. Therefore, the value of
$\varepsilon_{GS}$ should not depend greatly on the parameter
mistuning (compare the values of $\varepsilon_{GS}$ for the
$\omega_d=0.99$ (Fig.~1,\textit{c}) and $\omega_d=1.3$
(Fig.~1,\textit{d})). This conclusion agrees well with numerical
results of \cite{Zhigang:2000_GSversusPS}.

Let us briefly discuss why the onset of GS does not coincide with
any bifurcation point of the modified system (compare
Fig.~1,\,\textit{a,b} with Fig.~1,\,\textit{c,d}). The cause of
this non-coincidence is the influence of the external signal on
the modified system. As it has already been discussed above the
external signal (even if it is harmonic) can excite proper chaotic
oscillations in the dynamical system with periodic dynamics.
Therefore, the bifurcation points of the modified system under the
external signal will be shifted in the direction of the large
values of the $\varepsilon$-parameter in comparison with the
autonomous dynamics of the modified system. It is clear that the
onset $\varepsilon_{GS}$ of GS can not coincide with bifurcation
point of the autonomous modified system.

This statement is illustrated in Fig.~1,\,\textit{e} where the
bifurcation diagram for the response
system~(\ref{eq:ModRoesslers}) under the external harmonic signal
simulating the drive system signal is shown. One can see that all
bifurcation points of the modified system in the non-autonomous
regime are shifted in the direction of the larger values of
$\varepsilon$-parameter (compare Fig.~1,\,\textit{a} and
Fig.~1,\,\textit{e}).

The same effects take place when GS is observed in the discrete
maps. For example, GS takes place for the coupling parameter
values $\varepsilon\geq\varepsilon_{GS}\approx 0.32$ (see
\cite{Pyragas:1996_WeakAndStrongSynchro} for detail) in two
unidirectional coupled logistic maps
\begin{equation}
\begin{array}{l}
x_{n+1}=f(x_n),\\
y_{n+1}=f(y_n)+\varepsilon(f(x_n)-f(y_n)),
\end{array}
\label{eq:PyragasLogMap}
\end{equation}
where $f(x)=4x(1-x)$. Following the concept described above one
can construct the modified system
\begin{equation}
z_{n+1}=(1-\varepsilon)f(z_n)=az_n(1-z_n) \label{eq:ModLogMap}
\end{equation}
(where $a=4(1-\varepsilon)$) and obtain that the value
$\varepsilon_{GS}$ corresponds to the value  of $a\approx 2.72$ of
the equation (\ref{eq:ModLogMap}). For such $a$ value parameter
an attractor of the logistic map is the stable fixed point
${x^0=(a-1)/a}$.

Let us briefly discuss now the case of GS between oscillators of
different types or when the coupling between oscillators is not
diffusion. Several examples of such systems are
known (see, e.g., \cite{Pyragas:1996_WeakAndStrongSynchro,%
Rulkov:1996_AuxiliarySystem}). Obviously, if the coupling type is
diffusion the difference of the system types does not matter and
all reasons mentioned above remain true. But what happens when GS
takes place in the systems coupled in different way, rather than
in (\ref{eq:Oscillators1})? One of the examples of such systems
(see~\cite{Pyragas:1996_WeakAndStrongSynchro} for detail) is the
coupled R\"ossler (drive)
\begin{equation}
\begin{array}{l}
\dot x_{d}=-\alpha(y_{d}+z_{d}),\\
\dot y_{d}=\alpha(x_{d}+ay_{d},\\
\dot z_{d}=\alpha(p+z_{d}(x_{d}-c))
\end{array}
\label{eq:PyragasRoessler}
\end{equation}
($\alpha=6$, $a=0.2$, $p=0.2$, $c=5.7$) and Lorenz (response)
\begin{equation}
\begin{array}{l}
\dot x_{r}=\sigma(y_{r}-x_{r}),\\
\dot y_{r}=rx_{r}-y_{r}-x_{r}z_{r}+\varepsilon y_d,\\
\dot z_{r}=-bz_{r}+x_{r}y_{r}
\end{array}
\label{eq:PyragasLorenz}
\end{equation}
systems, where $\sigma=10$, $r=28$, $b=8/3$. It is
known~\cite{Pyragas:1996_WeakAndStrongSynchro}, that the value of
the coupling strength corresponding to the onset of GS is
$\varepsilon_{GS}\approx 6.66$. The amplitude of oscillations of
the $y_d$-coordinate of the R\"ossler system for selected
parameter values is about 10, the amplitude of the
$y_r$-coordinate of the Lorenz system being in autonomous regime
($\varepsilon=0$) is about 20. Obviously, the amplitude of the
external signal $\varepsilon_{GS}y_d$ introduced into the response
system (near the threshold of GS regime arising) is about 60. So,
the magnitude of the external force exceeds the level of proper
system oscillations in several times. This situation is
illustrated in Fig.~2 where the time series of $y_r(t)$
corresponding to the autonomous dynamics of the response
system~(\ref{eq:PyragasLorenz}) and the external force
$\varepsilon_{GS}y_d(t)$ are shown. It is clear, that the great
external force destroys completely proper dynamics of the response
system, the phase trajectory of the Lorenz system is moved into
the regions of the phase space with the strong dissipation and the
mechanism discussed above causes the appearance of GS.

\begin{figure}[tb]
\centerline{\includegraphics*[scale=0.39]{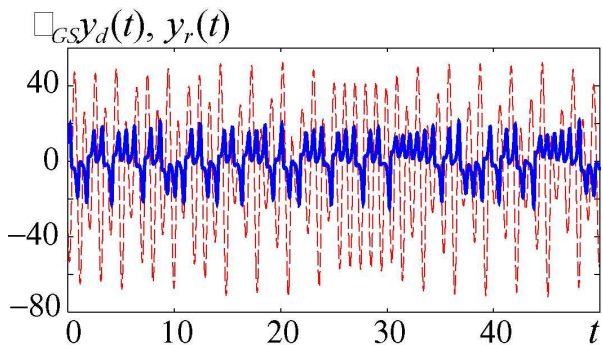}} \caption{Time
realization $y_r(t)$ corresponding to the autonomous dynamics of
the Lorenz system (solid line) and $\varepsilon_{GS}y_d(t)$
(dashed line) corresponding to the external signal introduced into
the response system near the onset of GS}
\end{figure}

In conclusion, we have explained GS appearance. The modified
system approach has been proposed to demonstrate the reasons of GS
arising. We have shown that the behavior of the response chaotic
system is equal to the dynamics of the modified system (with the
additional dissipation) under the external chaotic force. The
coupling parameter increase is equivalent to the simultaneous
growth of the dissipation and the amplitude of the external
signal.



We thank Anastasiya E. Khramova, Olga I. Moskalenko and Alexander
A. Tyschenko for the help in numerical experiments. We thank also
Svetlana V. Eremina for English support. This work has been
supported by U.S.~Civilian Research \& Development Foundation for
the Independent States of the Former Soviet Union (CRDF, grant
{REC--006}), Russian Foundation of Basic Research (project
05--02--16273), the Supporting program of leading Russian
scientific schools (project NSch-1250.2003.2) and the Scientific
Program ``Universities of Russia'' (project UR.01.01.371). We also
thank ``Dynastiya'' Foundation.




\begin{thebibliography}{17}
\expandafter\ifx\csname
natexlab\endcsname\relax\def\natexlab#1{#1}\fi
\expandafter\ifx\csname bibnamefont\endcsname\relax
  \def\bibnamefont#1{#1}\fi
\expandafter\ifx\csname bibfnamefont\endcsname\relax
  \def\bibfnamefont#1{#1}\fi
\expandafter\ifx\csname citenamefont\endcsname\relax
  \def\citenamefont#1{#1}\fi
\expandafter\ifx\csname url\endcsname\relax
  \def\url#1{\texttt{#1}}\fi
\expandafter\ifx\csname
urlprefix\endcsname\relax\def\urlprefix{URL }\fi
\providecommand{\bibinfo}[2]{#2}
\providecommand{\eprint}[2][]{\url{#2}}

\bibitem[{\citenamefont{{A. Pikovsky, M. Rosenblum,
J. Kurths}}(2001)}]{Pikovsky:2002_SynhroBook}
\bibinfo{author}{\bibnamefont{{A. Pikovsky, M. Rosenblum, J. Kurths}}},
  \emph{\bibinfo{title}{Synchronization: a universal concept in nonlinear
  sciences}} (\bibinfo{publisher}{Cambridge University Press},
  \bibinfo{year}{2001}).

\bibitem[{\citenamefont{{K. Murali, M. Lakshmanan}}(1994)}]{Murali:1993_SignalTransmission}
\bibinfo{author}{\bibnamefont{{K. Murali, M. Lakshmanan}}},
  \bibinfo{journal}{Phys. Rev. E} \textbf{\bibinfo{volume}{48}},
  \bibinfo{pages}{R1624} (\bibinfo{year}{1994}).

\bibitem[{\citenamefont{{T. Yang, C.W. Wu and L.O. Chua}}(1997)}]{Chua:1997_Criptography}
\bibinfo{author}{\bibnamefont{{T. Yang, C.W. Wu and L.O. Chua}}},
  \bibinfo{journal}{IEEE Trans. Circuits and Syst.}
  \textbf{\bibinfo{volume}{44}}, \bibinfo{pages}{469} (\bibinfo{year}{1997}).

\bibitem[{\citenamefont{{L. Glass}}(2001)}]{Glass:2001_SynchroBio}
\bibinfo{author}{\bibnamefont{{L. Glass}}}, \bibinfo{journal}{Nature}
  \textbf{\bibinfo{volume}{410}} (\bibinfo{year}{2001}).


\bibitem[{\citenamefont{{P. Paoli, A. Politi, R. Badii}}(1989)}]{Paoli:1989_ScalingBehavior}
\bibinfo{author}{\bibnamefont{{P. Paoli, A. Politi, R. Badii}}},
\bibinfo{journal}{Physica D}
  \textbf{\bibinfo{volume}{36}}, \bibinfo{pages}{263} (\bibinfo{year}{1989}).

\bibitem[{\citenamefont{{N.F. Rulkov, M.M. Sushchik, L.S. Tsimring,
H.D.I. Abarbanel}}(1995)}]{Rulkov:1995_GeneralSynchro}
\bibinfo{author}{\bibnamefont{{N.F. Rulkov, M.M. Sushchik, L.S. Tsimring,
H.D.I. Abarbanel}}}, \bibinfo{journal}{Phys. Rev. E}
  \textbf{\bibinfo{volume}{51}}, \bibinfo{pages}{980} (\bibinfo{year}{1995}).

\bibitem[{\citenamefont{{M.G. Rosenblum, A.S. Pikovsky,
J. Kurths}}(1997)}]{Rosenblum:1997_LagSynchro}
\bibinfo{author}{\bibnamefont{{M.G. Rosenblum, A.S. Pikovsky, J. Kurths}}},
  \bibinfo{journal}{Phys. Rev. Lett.} \textbf{\bibinfo{volume}{78}},
  \bibinfo{pages}{4193} (\bibinfo{year}{1997}).

\bibitem[{\citenamefont{{L.M. Pecora, T.L. Carroll}}(1990)}]{Pecora:1990_ChaosSynchro}
\bibinfo{author}{\bibnamefont{{L.M. Pecora, T.L. Carroll}}},
  \bibinfo{journal}{Phys. Rev. Lett.} \textbf{\bibinfo{volume}{64}},
  \bibinfo{pages}{821} (\bibinfo{year}{1990}).

\bibitem[{\citenamefont{{R. Brown,
L. Kocarev}}(2000)}]{Brown:2000_ChaosSynchro}
\bibinfo{author}{\bibnamefont{{R. Brown, L. Kocarev}}},
  \bibinfo{journal}{Chaos} \textbf{\bibinfo{volume}{10}}, \bibinfo{pages}{344}
  (\bibinfo{year}{2000}).

\bibitem[{\citenamefont{{S. Boccaletti, J. Kurths, G. Osipov, D.L. Valladares,
C.S. Zhou}}(2002)}]{Boccaletti:2002_SynchroPhysReport}
\bibinfo{author}{\bibnamefont{{S. Boccaletti, J. Kurths, G. Osipov, D.L. Valladares,
C.S. Zhou}}}, \bibinfo{journal}{Physics Reports}
  \textbf{\bibinfo{volume}{366}}, \bibinfo{pages}{1} (\bibinfo{year}{2002}).

\bibitem[{\citenamefont{{S. Boccaletti, L.M. Pecora, A. Pelaez}}(2001)}]{Boccaletti:2001_UnifingSynchro}
\bibinfo{author}{\bibnamefont{{S. Boccaletti, L.M. Pecora, A. Pelaez}}},
  \bibinfo{journal}{Phys. Rev. E} \textbf{\bibinfo{volume}{63}},
  \bibinfo{pages}{066219} (\bibinfo{year}{2001}).

\bibitem[{\citenamefont{{A.E. Hramov, A.A. Koronovskii}}(2004)}]{Hramov:2004_Chaos}
\bibinfo{author}{\bibnamefont{{A.E. Hramov, A.A. Koronovskii}}},
  \bibinfo{journal}{Chaos} \textbf{\bibinfo{volume}{14}}, \bibinfo{pages}{603}
  (\bibinfo{year}{2004}).

\bibitem[{\citenamefont{{K. Pyragas}}(1996)}]{Pyragas:1996_WeakAndStrongSynchr%
o}
\bibinfo{author}{\bibnamefont{{K. Pyragas}}}, \bibinfo{journal}{Phys. Rev. E}
  \textbf{\bibinfo{volume}{54}}, \bibinfo{pages}{R4508} (\bibinfo{year}{1996}).

\bibitem[{\citenamefont{{H.D.I. Abarbanel, N.F. Rulkov, M.M. Sushchik}}(1996)}]{Rulkov:1996_AuxiliarySystem}
\bibinfo{author}{\bibnamefont{{H.D.I. Abarbanel, N.F. Rulkov, M.M. Sushchik}}},
  \bibinfo{journal}{Phys. Rev. E} \textbf{\bibinfo{volume}{53}},
  \bibinfo{pages}{4528} (\bibinfo{year}{1996}).

\bibitem[{\citenamefont{{L.M. Pecora, T.L. Carroll, J.F. Heagy}}(1995)}]{Pecora:1995_statistics}
\bibinfo{author}{\bibnamefont{{L.M. Pecora, T.L. Carroll, J.F. Heagy}}},
  \bibinfo{journal}{Phys. Rev. E} \textbf{\bibinfo{volume}{52}},
  \bibinfo{pages}{3420–} (\bibinfo{year}{1995}).

\bibitem[{\citenamefont{{L.M. Pecora, T.L. Carroll}}(1991)}]{Pecora:1991_ChaosSynchro}
\bibinfo{author}{\bibnamefont{{L.M. Pecora, T.L. Carroll}}},
  \bibinfo{journal}{Phys. Rev. A} \textbf{\bibinfo{volume}{44}},
  \bibinfo{pages}{2374} (\bibinfo{year}{1991}).

\bibitem[{\citenamefont{{K. Pyragas}}(1997)}]{Pyragas:1997_CLEsFromTimeSeries}
\bibinfo{author}{\bibnamefont{{K. Pyragas}}}, \bibinfo{journal}{Phys. Rev. E}
  \textbf{\bibinfo{volume}{56}}, \bibinfo{pages}{5183} (\bibinfo{year}{1997}).

\bibitem[{\citenamefont{{Z. Zheng, G. Hu}}(2000)}]{Zhigang:2000_GSversusPS}
\bibinfo{author}{\bibnamefont{{Z. Zheng, G. Hu}}}, \bibinfo{journal}{Phys. Rev.
  E} \textbf{\bibinfo{volume}{62}}, \bibinfo{pages}{7882}
  (\bibinfo{year}{2000}).

\end{thebibliography}

\end{document}